%% file: article.tex
\documentclass{acm_proc_article-sp}

\usepackage{graphicx} 
\usepackage[hyphens]{url}

\usepackage{tabularx}
\newcolumntype{Y}{>{\centering\arraybackslash}X} 

\usepackage{multirow}
\usepackage{tikz} 
\usetikzlibrary{decorations.text}
\usetikzlibrary{calc,trees,positioning,arrows,chains,shapes.geometric,%
    decorations.pathreplacing,decorations.pathmorphing,shapes,%
    matrix,shapes.symbols}
\usepackage{verbatim}
\usepackage[inline]{enumitem}

\def\pdeclared{$P_{d}(app)$}
\def\pinferred{$P_{ifrd}(app)$}
\def\prequired{$P_{req}(app)$}

\newenvironment{definition}{}{}

\begin{document}

\title{Automatically Securing Permission-Based Software by Reducing the Attack Surface: An Application to Android} 

%
%
%
%
%

\numberofauthors{2} 
%
\author{
%
%
\alignauthor
Alexandre Bartel, Jacques Klein, Yves Le Traon\\
       \affaddr{University of Luxembourg, SnT}\\
       \affaddr{Luxembourg, Luxembourg}\\
       \email{firstName.lastName@uni.lu}
\alignauthor Martin Monperrus\\
       \affaddr{University of Lille}\\
       \affaddr{INRIA}\\
       \affaddr{Lille, France}\\
       \email{martin.monperrus@univ-lille1.fr}
\and  
}
\date{25 February 2012}

\maketitle
\begin{abstract}
\input{abstract}
\end{abstract}



\section{Introduction}
	\input{introduction}

\section{The Permission Gap Problem}\label{manifest-difficult}
	\input{writing-manifest}

\section{Analyzing Permissions}\label{manifest}
	\input{manifest-inference}

\section{Overview of Android}\label{android}
	\input{android-overview}

\section{Static Analysis for Android}\label{application-on-android}
	\input{tool-design-and-implementation}

\section{Evaluation}\label{empirical-study}
	\input{experimentations}

\section{Related Work}\label{related-work}
	\input{related-work}

\section{Conclusions and Perspectives}\label{conclusion}
	\input{conclusion}


%
\bibliographystyle{abbrv}
\bibliography{bib/bib}  
%
%
\balancecolumns

\end{document}

%% file: abstract.tex
A common security architecture, called the permission-based security model (used e.g. in Android and Blackberry), entails intrinsic risks. For instance, applications can be granted more permissions than
they actually need, what we call a ``permission gap''. Malware can
leverage the unused permissions for achieving their malicious goals,
for instance using code injection. In this paper, we present an
approach to detecting permission gaps using static analysis. Our
prototype implementation in the context of Android shows that the
static analysis must take into account a significant amount of
platform-specific knowledge. Using our tool on two datasets of Android
applications, we found out that a non negligible part of applications
suffers from permission gaps, i.e. does not use all the permissions
they declare. 

%% file: introduction.tex
Android is one of the most widespread mobile operating system in the world accounting 52\% market share  
\cite{Gartner2011}. 
More than 300 000 Android applications available on dozens of application markets can be installed by end users. 
The other side of the coin is that all kinds of malware are waiting to be installed on thousands of Android devices. For instance, Zeus \cite{Hoffman2011} sends banking information to malicious servers. This motivates researchers and engineers to devise security models, architectures and tools that are able to mitigate the malware harmfulness.

The security architecture of Android, the Google Chrome browser
extension system and the Blackberry platform, all use a similar security model called the permission-based security model \cite{DBLP:conf/ccs/BarreraKOS10}. 
A permission-based security model can be loosely defined as a model in which
1) each application is associated with a set of permissions that allows accessing certain resources\footnote{Contrary to the traditional Unix permission system where permissions are at the level of users, not applications.};
2) permissions are explicitly accepted by users during the installation process and
3) permissions are checked at runtime when resources are requested.

This permission model entails intrinsic risks. For instance, not all users may be able to cleverly reject powerful permissions at installation time. Malwares may also use platform vulnerabilities to circumvent runtime permission checks. Finally, applications can be granted more permissions than they actually need, what we call a ``permission gap''.
Malwares can leverage the unused permissions for achieving their malicious goals and have many ways to do so, for instance using code injection or return-oriented programming \cite{Davi2011}. Identifying permission gaps means reducing the risks for an application to be compromised, also known as reducing the application attack surface \cite{Manadhata2011}.\newline

Let us make an analogy with a firewall. In a correctly configured
firewall only the ports that are used are open. All the other ports are
closed. However if the firewall is misconfigured, some unused ports remain open and the attack surface of the infrastructure behind the firewall is larger.
For instance, let us assume that an information system internally uses a remote
shell service on port 544. If port 544 is open on the firewall, an attacker could perform attacks on the remote shell server located behind the firewall.
In the same way, an application that requires too many permissions, i.e. that suffers from a permission gap, may allow an attacker who compromised the application to access more resources than he should have.

Permission gaps appear because the process of declaring application permissions is manual and error-prone:
Android framework developers manually document which permissions are required for each system resource,
and Android application developers manually declare the permissions they \emph{think} are needed.
This paper presents an approach to support those \emph{manual} software engineering activities with an \emph{automated} tool.
This approach secures permission-based software in the sense that it reduces the attack risks (not in the sense that the resulting software is unattackable).

Our tool, called COPES (COrrect PErmissions Set), proceeds as follows.
First, using static analysis, it extracts from the Android framework bytecode a table that  maps every method of the API to a set of permissions the method needs to be executed properly.
Second, COPES lists all framework methods used by an application, based on static analysis of the application bytecode.
Third, COPES computes the set of permissions that are required for the application to run, which means that all permissions in this set are may be at least once used in the application, and consequently no permission gap remains.
Eventually, COPES computes the permission gap as the difference between the declared permissions and the required permission.
By listing the permission checks per framework method, COPES can also help Android designers to comprehensively document the framework.

To sum up, the contribution of this paper is an approach to identify and fix permission gaps in permission based software. More specifically:\newline
$\bullet$ We show that the permission-based security model can be expressed within a boolean matrix algebra. This algebra is not specific to Android.\newline
$\bullet$ We present a novel methodology to compute a close approximation of the required permission set and the permission gap based on static analysis, as opposed to concurrent work that uses testing \cite{Felt2011a}.\newline
$\bullet$ We discuss the design and the implementation of the approach for the Android platform.\newline
$\bullet$ We evaluate our approach on 742+679 Android applications and we show that 94+35 applications suffer from a permission gap.

The reminder of this paper is organized as follows. In Section \ref{manifest-difficult} we explain
why reducing the attack surface is important and present a short study supporting our intuition.
In Section \ref{manifest} we propose a formalization for permission-based software and a 
generic method for deriving correct application permission sets. 
In Section \ref{android} we describe the Android system and its access control mechanisms.
Then, in Section \ref{application-on-android} we apply the generic method 
on the Android system. Experiments we conducted and results are presented and discussed in Section \ref{empirical-study}.
We present the related work in Section \ref{related-work}.
Finally we conclude the paper and discuss open research challenges in Section \ref{conclusion}.

%% file: writing-manifest.tex

This section further details the permission gap problem introduced in Section 1, and presents short empirical facts showing that this problem actually happens in practice.

\subsection{Possible Consequence of a Permission Gap}

Let us consider an Android application, $app_{wrong}$, that is able to communicate with external servers since it is granted the INTERNET permission.
Moreover, $app_{wrong}$ has declared permission CAMERA while not using it.
The CAMERA permission allows the application to take picture without user intervention, i.e. the permission gap consists of one permission: CAMERA.
Unfortunately, $app_{wrong}$ uses a native library on which a buffer-overflow exploit has recently been discovered. As a result, through specific payloads, attackers are able to attack devices that are running $app_{wrong}$ in order to take pictures using the device's camera and send them to a remote location on the Internet.

On the contrary, if $app_{wrong}$ did not declare CAMERA, this attack would not have been possible, and the consequences of the buffer-overflow exploit would have been mitigated.
As noted by Manadhata \cite{Manadhata2011}, reducing the attack surface does not mean no risks, but less risks.
In order to show that this example of misconfigured application is not artificial, we now discuss a short empirical study on the declaration of two permissions on 1000+ Android applications.

\subsection{Declaration and Usage of Permissions \\ CAMERA and RECORD\_AUDIO}

We conducted a short empirical study on a 1000+ Android applications downloaded from the Freewarelovers application market\footnote{\url{http://www.freewarelovers.com/android/}}. 
For permissions CAMERA and RE\-CORD\_AU\-DIO, we grepped the source code of the Android framework to approximate the set of methods requiring one of them. These two sets of methods are noted $M_{CAMERA}$ and $M_{RECORD\_AUDIO}$.
Then, we computed the list $A$ of all the applications which declare CAMERA or  RECORD\_AUDIO. Next, we took each application $app$ $\in$ $A$ individually and we checked the application uses at least one method of $M_{CAMERA}$ and $M_{RECORD\_AUDIO}$ by analyzing the application's bytecode.  If it is not the case, it meant that $app$ is not using the corresponding permission. 
When this happened, we modified the application manifest that declares the permission and run the application again to make sure that our grepping approximation did not yield false positives.

There are 7/82 applications that declare CAMERA while not using it.
Similarly, 3/35 applications declare but do not use RECORD\_AUDIO .
Those results confirm our intuition: declared permission lists are not always
required, and permission gaps indeed exist. 
Developers would benefit from a tool that automatically infers the set of required permissions and computes the permission gap.

%% file: manifest-inference.tex

In this Section we formalize the concept of permission-based software and propose a generic methodology to compute a mapping from code to permissions that are required for the application to run. 

Permission-based software is conceptually divided in three layers:
1) the core platform which is able to access all system resources (e.g. change the network policy), it is generally the operating system;
2) a middleware responsible for providing a clean application programming interface (API) to the OS resources and for checking that applications have the right permissions when they want accessing them;
3) applications built on top of the middleware. They have to explicitly declare the permissions they require.
Layers \#2 and \#3 motivate the generic label ``permission-based software''.
Since the middleware also hides the OS complexity and provides an API, it is sometimes called, as in the case of Android, a ``framework''.

Let us now discuss those terms more in depth and then show how to infer the list of permissions required by a permission-based application.

\subsection{Definitions}
\label{subsec:def}

	\begin{definition}\textbf{Framework}
	\label{def:framework}  
	A framework  $\mathcal{F}$ is a layer that enables applications to access resources available on the platform. 
	We model it as a bi-partite graph
between framework API methods and resources.
	\end{definition}\newline 
 \emph{Example}: In the case of Android, $\mathcal{F}$ is the Android $2.2$ Java Framework
composed of 4071 classes and 126660 methods.
To access a resource, an Android application has to make a method call that goes through $\mathcal{F}$.

	\begin{definition}\textbf{Permission-based system}
	\label{def:permission-based-system}
	A permission-based system is composed of  at least one framework, a list of permissions and a list of protected resources. 
Each protected  resource is associated with a fixed list of permissions
	\end{definition}

	\begin{definition}\textbf{Entry point}
	\label{def:entry-point}
	An entry point of a framework is a method that an application 
	can use (e.g. public or documented). Constructors are also considered as entry points.
	We denote $Entry_{\mathcal{F}}$ the set of all entry points of $\mathcal{F}$.
	\end{definition}\newline
\emph{Example}: One of the entry points of the Android framework is the method \textit{getAccounts()}
from class \textit{AccountManager}.

An application can call any public method of the framework. 
Some methods accessing some system resources (like an account) are protected by one or more permissions.
Let us suppose that the method \textit{getAccounts()} allows access to a set of accounts and is protected by one permission \texttt{GET\_ACCOUNTS}.
An application can successfully call method \textit{getAccounts()} if and  only if it declared \texttt{GET\_ACCOUNTS} in the application-specific list (this list is contained in a ``manifest'', we shall use this term later in the paper).

	\begin{definition}\textbf{Permission}
	\label{def:pep}
	A permission is a token that an application needs to access a specific resource. 
We make no assumptions on permissions, and we consider them as independent (neither grouped, nor hierarchical) .
	\end{definition}\newline
\emph{Example}: Developers of an Android application define a list of permissions in a file called the Manifest. 
To read contact information, the manifest of the application must declare the \texttt{READ\_CONTACT} permission. 

Permissions can be checked at different levels in the system. We call high-level permissions the set $P = \{p_1, p_2, ..., p_n\}$ of permissions that are checked at the framework level. Low-level permissions are permissions that are checked at the operating system level.

	\begin{definition}\textbf{High-level permission}
	\label{def:high-level-permission}
	A high-level permission, is a permission that is only checked at the framework level.
	\end{definition}\newline
\emph{Example}: In the case of Android, \texttt{READ\_CONTACT} is a high-level permission. 

	\begin{definition}\textbf{Low-level permission}
	\label{def:low-level-permission}
	A low-level permission is a permission associated with a high-level permission and is checked at a lower level than
the framework level.
	\end{definition}\newline
\emph{Example}:  
They are 115 permissions in the Android system, while 8 permissions are checked at a low-level. This shows that the framework is responsible for much of the work related to permissions. Note that if a permission is checked at the operating system level, it is not possible to detect that an application uses it by only analyzing the framework.

	\begin{definition}\textbf{Declared permission}
	\label{def:declared-permission}
	A declared permission for an application $app$ is a permission which is in the permission list of $app$.
	The set of all declared permission for an application $app$ is noted \pdeclared.
	\end{definition}

	\begin{definition}\textbf{Required permission}
	\label{def:required-permission}
	A required permission for an application $app$ is a permission associated with a resource that $app$ uses at least once.
	The set of all required permissions for an application $app$ is noted \prequired.
	\end{definition}\newline
\emph{Example}: For an application $app$, if the set of required permissions \prequired~is equal to the set of declared permissions \pdeclared, the permission attack surface is minimal.

	\begin{definition}\textbf{Inferred permission}
	\label{def:inferred-permission}
	An inferred permission for an application $app$  is a permission that an analysis technique found to be required for $app$.
	This paper presents such a technique and computes a set of inferred permissions noted \pinferred. 
	Depending on the analysis technique used, the inferred permission list may be either an over- or an under- approximation of
	the required permission list. 
	When using static analysis techniques, the inferred permission list may be an \emph{over approximation} (\prequired~$\subseteq$ \pinferred). 
	The inferred permission list may be an \emph{under-approximation} of the required permission list (\pinferred~$\subseteq$ \prequired) when using testing techniques (testing-based analysis observes only the executed permissions and potentially misses some permission checks depending on the completeness of the input data).
	\end{definition}

When developers write manifests, they write \pdeclared~by trying to guess \prequired~based on documentation and trial-and-errors.
In this paper, we propose to automatically infer a permission list \pinferred~in order to avoid this manual and error-prone activity.
We take a special care in minimizing the difference between \pinferred~and \prequired.

\subsection{A Calculus for Permissions}
\label{subsec:permCalculus}

This section describes the permission gap inference as a clean and regular calculus on top of boolean matrix algebra.
More importantly, while permission inference is at heart a reachability analysis (does the application reach a permission check?), this calculus factorizes much of the static analysis, hence is much more efficient.

	Let $app$ be an application. The \emph{access vector} for $app$ is a boolean vector $AV_{app}$ representing the entry points of the framework reachable from $app$.  Thus, the length of vector $AV$ is the number of entry points of  $\mathcal{F}$. 
	An element of the vector is set to \emph{true} if the corresponding entry point is called by the application. Otherwise it is set to \emph{false}.
Let us consider a framework with four entry points ($e_1$, $e_2$, $e_3$, $e_4$), and an application $app$ with the following access vector, expressing that $app$'s code may reach $e_1$, $e_2$ and $e_3$ but not $e_4$:
$$ AV_{app} = \left(1, 1, 1, 0\right) $$ 
We define the \emph{permission access matrix} $M$ as a boolean matrix which represents the relation between entry points of the
	framework and permissions. Rows represent entry points of the framework and columns 
	represent permissions. 
	A cell $M_{i,j}$ is set to \emph{true} if the corresponding entry point (at row $i$)  
	accesses a resource protected by the permission represented by column $j$. Otherwise it is set to \emph{false}.
For a framework with four entry points ($e_1$, $e_2$, $e_3$ and $e_4$) 
and three permissions ($p_1$, $p_2$ and $p_3$), the permission access matrix could be:
$$ M = 
\bordermatrix{
  & p_1	& p_2 & p_3  \cr
e_1 & 1 & 0 & 0  \cr
e_2 & 1 & 0 & 0  \cr
e_3 & 0 & 0 & 0  \cr
e_4 & 0 & 1 & 0  \cr
}
$$
This means that $e_1$ and $e_2$ require permission $p_1$, $e_3$ requires no permission and that $e_4$ requires permission $p_2$.

	Let $app$ and $\mathcal{F}$ be an application and a framework respectively.  
	The inferred permissions vector, $IP_{app}$, is a boolean vector representing the set of inferred permissions for application $app$.
	We have $IP_{app} = AV_{app} \times M$ by using the boolean operators AND and OR instead of arithmetic multiplication and addition in the matrix calculus.
	A cell $IP_{app}(k)$ is set to \emph{true} if the permission at index $k$ is required by $app$. Otherwise it is set to \emph{false}.
Note that \pinferred~is the set of all permissions set to \emph{true} in $IP_{app}$, i.e. \pinferred $ = \{ permission_x | IP_{app}(x) \}$. 
Using $AV_{app}$ and $M$ from the two previous examples, the inferred permissions vector for $app$ is:
\begin{eqnarray*}
IP_{app} & = &  \left( \begin{array}{ccccc} 1 & 1 & 1 &  0 \end{array}\right) \cdot
\left( \begin{array}{ccc}
1 & 0 & 0 \\
1 & 0 & 0 \\
0 & 0 & 0 \\
0 & 1 & 0 \end{array} \right) \\
IP_{app} & = & \left( \begin{array}{ccc} 1 & 0 & 0  \end{array}\right)
\end{eqnarray*}
Application $app$ should declare permissions $p_1$.

\subsection{Extraction of $M$ and $AV$}
In this section we present a methodology to extract the permission access matrix $M$ of a framework $\mathcal{F}$. 
This methodology is based on a static analysis of the framework $\mathcal{F}$. 
Our idea is to first to compute a call graph for every entry point of the framework and then to detect whether or not permission checks are present in the call graph. A call graph is a directed graph G containing a set of vertices V representing method calls and a set of arcs A representing links between method calls.

A permission enforcement point (PEP) is a vertex of a 
	call graph whose signature corresponds to a system 
 method which checks permission(s). Each PEP is associated with a list of required permissions $perms_{PEP}$.
In Figure \ref{fig:generic-application-framework}, the call graph starting from entry point $e_4$ reaches $ck_2$,  
a call to a PEP. 
To localize in which methods PEP are called, we traverse a call graph $G = (V,A)$ generated  from the framework and check whether a vertex $V_{PEP}$ is a PEP. 
Methods which directly check for permissions are represented as vertice $V_{M_i}$ ($i \in$ \{1,2,...,k\}), such
that $(V_{M_i}, V_{PEP}) \in A$.

	We compute one call graph $G_i$ per entry point $e_i$ of the framework ($i \in$ \{1,2,...,n\})\footnote{This is especially important for field-sensitive static analyzes)}.
Then , matrix $M$ is constructed as follows:
$M$ is set as a matrix of size  (|entry points| $\times$ |high level permissions|);
all elements of $M$ are initialized to false;
for each $e_i$ that reaches one ore more PEP,  and for each permission $j$ in $perms_{PEP}$, $M(i,j) = true$. 
In other terms, M is a condensed version of the reachability information that is latent in call graphs.
 For instance, a framework with four entry points $(e_1,e_2,e_3, e_4)$, and three permissions $(p_1,p_2,p_3)$ is presented in the lower part
of Figure \ref{fig:generic-application-framework}.
For every of those entry points a call graph is constructed. 
Three of those call graphs have a PEP node: $e_1$ and $e_2$ have PEP $ck_1$ which checks permission $p_1$ and $e_4$ has PEP $ck_2$ which checks permission $p_2$.
On the figure a dashed arrow connects each PEP to the permission(s) it checks.
The framework matrix, noted $M_{ex}$, is then the same as matrix $M$ (see Section \ref{subsec:permCalculus}).

Extracting $AV$ simply means listing the list of entry points of a framework $\mathcal{F}$ called by an application $app$.
The application example in Figure \ref{fig:generic-application-framework} features a single
entry point, $s$. From $s$ a call graph $G_{ex}$ is generated. All elements of vector $AV_{ex}$ of length $n=4$ are initially set to $false$:
$ AV_{ex} = \left(0, 0, 0, 0\right) $.
Then for every vertex of $G_{ex}$ which is a call to the example framework, the corresponding element of $AV_{ex}$ is set to $true$. 
In the example, there are three such vertices (represented as entry points $e_1$, $e_2$ and $e_3$ in Figure \ref{fig:generic-application-framework}). This
leads to the following vector
$ AV_{ex} = \left(1, 1, 1, 0\right) $.

\subsection{Computing the Permission Gap}
We compute the inferred permission vector according to the definition presented in Section 3.1. 
The inferred permission list corresponds to permissions set to \emph{true} in $IP_{app}$. 
In Figure \ref{fig:generic-application-framework}, using matrix $M_{ex}$ and vector $AV_{ex}$ generated above for the example framework and application, we obtain an list of inferred permissions only containing $p_1$.
The permission gap is the difference between the permissions extracted from $IP_{app}$ and the declared permissions \pdeclared.
If the application declares $p_1$ and $p_2$, the permission gap is $\{p_2\}$.

\begin{center}
\begin{figure}
\begin{center}
	\input{figures/generic-framework-application-graph}
\caption{\label{fig:generic-application-framework}Application and Framework Example}
\end{center}
\end{figure}
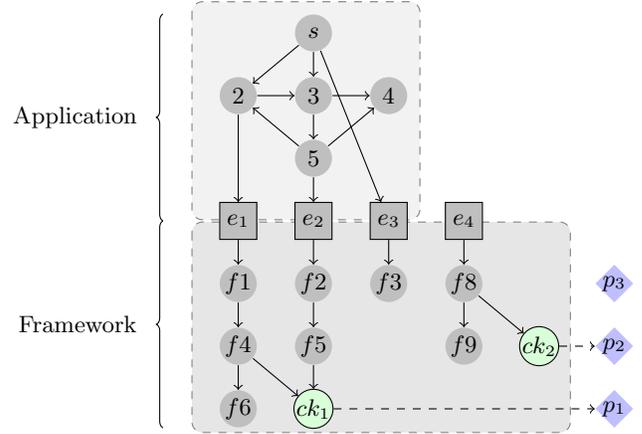
\end{center}

%% file: figures/generic-framework-application-graph.tex

\pgfdeclarelayer{background}
\pgfdeclarelayer{foreground}
\pgfsetlayers{background,main,foreground}
\tikzstyle{block} = [draw,fill=white!20,minimum width=4em]
\begin{tikzpicture}
  [node distance=.5cm,
  start chain=going below,]

		  \tikzstyle{vertex}=[circle,fill=black!25,minimum size=14pt,inner sep=0pt]
		  \tikzstyle{vInterface}=[block,fill=black!25,minimum size=14pt,inner sep=0pt]
		  \tikzstyle{vPermission}=[diamond,fill=blue!25,minimum size=14pt,inner sep=0pt]
  		\tikzstyle{tuborg}=[decorate]
  		\tikzstyle{tubnode}=[midway, right=2pt]
		  \foreach \name/\x/\y in {s/1/0, 2/0/1, 3/1/1, 4/2/1, 5/1/2}
		    \node[vertex] (G-\name) at (\x,-\y/1.2) {$\name$};
		  \node[vInterface] (G-e1) at (0,-3/1.2) {$e_{1}$};
		  \node[vInterface] (G-e2) at (1,-3/1.2) {$e_{2}$};
		  \node[vInterface] (G-e3) at (2,-3/1.2) {$e_{3}$};
		  \node[vInterface] (G-in) at (3,-3/1.2) {$e_{4}$};
		  \foreach \from/\to in {s/2, s/3, s/e3, 2/e1, 2/3, 3/5, 3/4, 5/2, 5/4, 5/e2}
		    \draw[->] (G-\from) -- (G-\to);

		  \foreach \name/\x/\y in {f1/0/4, f2/1/4, f3/2/4, f4/0/5, f5/1/5, f6/0/6, f8/3/4, f9/3/5}
		    \node[vertex] (G-\name) at (\x,-\y/1.2) {$\name$};
		  \foreach \name/\x/\y/\n in {ck/1/6/1, ck/4/5/2}
		    \node[draw,vertex,fill=green!15] (G-\name\n) at (\x,-\y/1.2) {$\name_\n$};
		  \foreach \from/\to in {e1/f1, e2/f2, e3/f3, f1/f4, f2/f5, f5/ck1, f4/f6, f4/ck1, f8/f9, f8/ck2, in/f8} 
		    \draw[->] (G-\from) -- (G-\to);

		  \node[vPermission] (G-p3) at (5,-4/1.2) {$p_{3}$};
		  \node[vPermission] (G-p2) at (5,-5/1.2) {$p_{2}$};
		  \node[vPermission] (G-p1) at (5,-6/1.2) {$p_{1}$};
		   \draw[->,dashed] (G-ck1) to (G-p1);
		   \draw[->,dashed] (G-ck2) to (G-p2);

		  \node[] (G-00) at (0,0) {};
		  \node[] (G-10) at (2,0) {};
		  \node[] (G-01) at (0,-3/1.2) {};
		  \node[] (G-11) at (2,-3/1.2) {};
		  
			\node[] (G-F00) at (0,-3/1.2) {};
		  \node[] (G-F10) at (4,-3/1.2) {};
		  \node[] (G-F01) at (0,-6/1.2) {};
		  \node[] (G-F11) at (4,-6/1.2) {};
\begin{pgfonlayer}{background}
        \path (G-00.west |- G-10.north)+(-0.5,0.3) node (a) {};
        \path (G-01.north -| G-11.east)+(+0.3,-0.1) node (b) {};
        \path[fill=black!5,rounded corners, draw=black!50, dashed]
            (a) rectangle (b);
        \path (G-F00.west |- G-F10.south)+(-0.5,0.1) node (a) {};
        \path (G-F01.south -| G-F11.east)+(+0.3,-0.2) node (b) {};
        \path[fill=black!10,rounded corners, draw=black!50, dashed]
            (a) rectangle (b);
\end{pgfonlayer}

			\draw[tuborg, decoration={brace}] let
			    \p1=(G-s.north), \p2=(G-e1) in
					    ($(-1,\y2)$) -- ($(-1,\y1)$) node[tubnode, left]  {Application$\hspace{.3cm}$};
			\draw[tuborg, decoration={brace}] let
			    \p1=(G-e1), \p2=(G-f6.south) in
					    ($(-1,\y2)$) -- ($(-1,\y1)$) node[tubnode, left]  {Framework$\hspace{.3cm}$};

\end{tikzpicture}

%% file: android-overview.tex

This section gives an overview of the architecture and specificity of the Android software stack. 
We detail how applications  access the framework$_{\mathcal{F}}$ and where access control is enforced with respect to  permissions.

\subsection{\label{sub:software-stack}Software Stack}
Android is a system with different layers.
It consists of a modified Linux kernel, C/C++ libraries, a virtual machine called Dalvik, a Java framework and a set of basic applications (including a phone application). 
Applications for Android are written in Java.
An Android application is packaged into a Android package file (ending in .apk) which contains the Dalvik 
bytecode, data (pictures, sounds, ...) and the Android manifest file. 
The developer defines permissions the application may use in this manifest. 

An Android application is made of \emph{components} which can be:
\begin{enumerate*}
	\item an \emph{Activity} which is a user interface;
	\item a \emph{Service} which runs in background;
	\item a \emph{BroadcastReceiver} which listens for ``intents'' (a kind of message comparable to inter processes communication, aka IPC);
	\item a \emph{ContentProvider} which is a kind of backend database used to store and share  data\footnote{An application uses URIs (Uniform Resource Identifiers from RFC \#2396) to locate and work with local or remote  content providers}.

\end{enumerate*}

\subsection{Services}\label{subsec:service}

Applications define services that can be used by other applications. 
They also communicate with the operating system using  a special kind of services called \emph{system services} that are used by the system for enforcing permission checks.
System services are specific services running in a specific scope (called the ``system server'') and allow applications to access system resources.
Those resources may be protected by Android permissions. 
The permission checks associated to services
are mostly implemented in Java, but Android 
also checks permissions in C++ services, content providers and when using
intents.
In this paper, we focus on the former (Java services), the impact of this focus
is discussed in Section \ref{empirical-study}.

\begin{figure}[h]
\begin{center}
	\input{figures/android-system-service}
\caption{Android System Service}
\label{fig:android-system-service}
\end{center}
\end{figure}
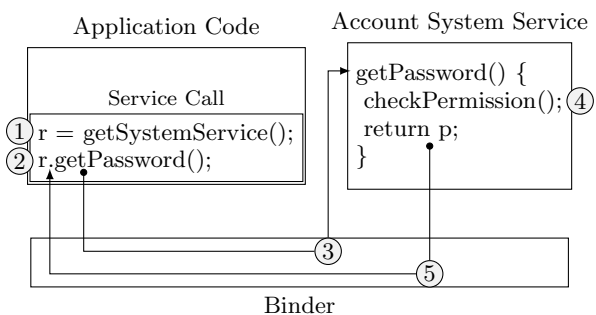

Applications synchronously communicate with others services (deployed from other applications or the OS) through a mechanism called  \emph{Binder}.
The first step to communicate with a remote service is to dynamically get a reference to the service by calling \texttt{Context.getSystemService()} (step 1 in Figure \ref{fig:android-system-service}). 
The next step is to call a method on the reference (step 2 in Figure \ref{fig:android-system-service}).  
A special component, called ``binder'' is responsible for delivering references, intercepting  and redirecting that service calls to the remote service which performs the actual computation (steps 3 and 5 in Figure \ref{fig:android-system-service}). The system service is responsible for enforcing the permission policy (step 4 in Figure \ref{fig:android-system-service}).

\subsection{Permissions and Application Installation}

When installing an Android application from an application market, 
the user has to approve as a whole (or reject as a whole) all the permissions the application declared in its manifest.
If all permissions are approved, the application is installed and mapped with the corresponding permissions. 
Moreover, it receives a device-specific user id (UID) 
and group memberships for the permissions that are mapped with Unix groups. 
For instance, an application Foo is given two group memberships \texttt{net\_bt} and \texttt{inet} when associated with permissions \texttt{BLUETOOTH} and \texttt{INTERNET}, respectively.
In other terms, the standard Unix ACL is used as an implementation means for checking permissions.

Android 2.2 declares 107 (115-8) high-level permissions, high-level in the sense that they are enforced at the framework level
(ex: to read contact information an application must have the  \texttt{READ\_CONTACT} permission). 
Note that high-level permissions are related to \emph{where} permissions are checked (in the framework) not \emph{how} (mostly using 
Unix group memberships by the Dalvik virtual machine).
There are eight high-level permissions that may also be indirectly enforced at the kernel level by checking unix group IDs
(ex: to create a socket an application has to have the \texttt{INTERNET} permission to be in the \texttt{inet} group).

In Section \ref{manifest}, we have defined a generic model and methodology to generate a matrix $M$ which maps entry points of a framework$_{\mathcal{F}}$ to
permissions. We have seen  in Section \ref{android} that the Android system fits in the model and contains a framework corresponding to $_{\mathcal{F}}$.
The next Section presents a static analysis to extract $M$ from the Android framework$_{\mathcal{F}}$ and to infer the list of required permissions (as opposed to declared permissions) for an Android application.

%% file: figures/android-system-service.tex

\pgfdeclarelayer{background}
\pgfdeclarelayer{foreground}
\pgfsetlayers{background,main,foreground}

\tikzstyle{sensor}=[draw, fill=blue!20, text width=5em, 
    text centered, minimum height=2.5em]
\tikzstyle{ann} = [above, text width=5em]
\tikzstyle{naveqs} = [sensor, text width=6em, fill=red!20, 
    minimum height=12em, rounded corners]

\tikzstyle{zygote} = [draw, fill=white!20, text width=6em, 
	text centered, minimum height=15em]

\tikzstyle{app} = [draw, fill=white!20, text width=8em, 
	text centered, minimum height=5.6em]
\tikzstyle{service} = [draw, fill=white!20, text width=8em, 
 minimum height=6em]
\tikzstyle{bindervoid} = [ minimum width=0.2em, 
	text centered, minimum height=6em]
\tikzstyle{binder} = [draw, fill=white!20, minimum width=22em, 
	text centered, minimum height=2em]
\tikzstyle{mystep} = [draw, fill=black!5, circle, inner sep=.1em]

\tikzstyle{systemserver} = [draw, fill=white!20, text width=6em, 
	minimum height=15em]
\tikzstyle{servicemanager} = [draw, fill=white!20, text width=6em, 
	text centered, minimum height=15em]
\tikzstyle{binderdriver} = [draw, fill=white!20, text width=20em, 
	text centered, minimum height=3em, minimum width=24em]
\tikzstyle{stub} = [draw, dashed, rotate=90, minimum height=1.4em, 
	text centered]
\tikzstyle{activity} = [draw, fill=white!20,
	minimum height=13.3em,  minimum width=5em]
\tikzstyle{proxy} = [draw, dashed, rotate=90, minimum height=1.2em, 
	minimum width=1em, text centered]
\def\blockdist{5em}
\def\edgedist{3em}

\begin{tikzpicture} [node distance = 6em]

		\draw node (app1) [app, text width = 10.7em, label=above:Application Code] at (0,0) {};
		\draw node (API1) [draw, fill=white!20, text width=10.5em, yshift=-1.3em, label=above:{\small Service Call}] {{\footnotesize r = getSystemService();} r.getPassword();};

		\draw node (bindervoid1) [right of=app1, bindervoid, minimum width=0em] {};
		\draw node (binder1) [xshift=-.55em, below of=bindervoid1, binder, label=below:{\footnotesize Binder}] {};

		\draw node (service1) [right of=bindervoid1, service, text width=8.5em, label=above:{\footnotesize Account System Service}] {getPassword() \{ \\
			{\ }checkPermission();\\
			{\ }return p;\\
			\}};


		\draw node (step1) [mystep] at (-1.95, -.2) {1};
		\draw node (step2) [mystep] at (-1.95, -.6) {2};

		\draw node (step3) [mystep] at (2.15, -1.80) {3};

		\draw node (step4) [mystep] at (5.5, .2) {4};

		\draw node (step5) [mystep] at (3.5, -2.1) {5};

		\coordinate (path1start) at (-1.1, -.75);
		\coordinate (path1end) at (2.45, .6);
		\draw node [circle, minimum width=.25em, draw, fill, color=black, inner sep =0em] at (path1start) {};
		\path [draw, -latex] (path1start) |- (step3) |- (path1end);
		\coordinate (path2end) at (-1.55, -.70);
		\coordinate (path2start) at (3.5, -.4);
		\draw node [circle, minimum width=.25em, draw, fill, color=black, inner sep =0em] at (path2start) {};
		\path [draw, -latex] (path2start) |- (step5) -| (path2end);
\end{tikzpicture}

%% file: tool-design-and-implementation.tex
Our approach to detecting permission gaps presented in Section \ref{manifest} is implemented with two tools.
One extracts from a permission-based framework a binary matrix that maps framework methods to permissions, we call it the \emph{mapper}.
The other extracts from application code the list of framework methods used, we call it the \emph{sniffer}. 
In COPES, both tools are based on static analysis.
Implementing both tools was much more difficult than expected. In other terms, there was a significant gap between the regularity and the conciseness of the approach presented in Section \ref{manifest} and the actual implementation.

The key insights of our analysis are related to correctly handling the service and binder mechanisms of Android (see \ref{subsec:service}).
This section presents our solutions to the most important issues in order to 1) enable other researchers to replicate our results, and 2) facilitate the implementation of the approach for another permission-based platform.

\subsection{Framework Call Graphs}
\label{sec:call-graph}

The core of our approach consists of building and manipulating call graphs.
COPES call graph construction leverages the Soot call graph analysis  Spark \cite{lhothend03spark} together with the service mapping information described in below in \ref{sec:binder} and \ref{sec:binder-identity}.

We run Spark in context\--insensitive, path\--insensitive, flow\--insensitive, field\--sensitive
mode  to generate the call graph. 
In context-insensitive mode, every call to a same method are merged to a single edge independently of the context (receiver and parameters values).
A path-insensitive analysis ignores conditional branching hence takes into account all paths of method bodies.
The call graph construction is flow-insensitive since it does not consider the order of executions of instructions.
It is also field-sensitive because it uses and propagates initialization data (e.g. constructor calls) to reduce the number of egdes.

Spark requires an entry point (usually a main) in order to apply its aggressive edge removal techniques. 
In the case of an API (such as the Android API), there is no ``main''. 
Hence, we build one call graph per public method of the Android API by creating one fake main method per public class of the framework (for Android, \texttt{android.*} and \texttt{com.android.*}). 
We can also build an artificial main calling all public methods, which is conceptually equivalent yet less scalable\footnote{we were not able to extract such a call graph on a machine with 24GB RAM}.

\subsection{Extracting Permission Enforcement Points}

Permission Enforcement Points in Android are method calls to certain method of classes \texttt{Con\-text} and \texttt{Con\-textWrap\-per} (for instance method \texttt{checkPermission}).
Those me\-thod calls can be resolved statically.
However, the actual permission(s) that are checked are dynamically set by a String parameter or sometimes, an array of strings. 
Thus, when a check permission system method is found in the call graph, a basic analysis is only able to tell that a permission check occurs, but not which precise permission.

To overcome this issue, we have implemented a String analysis as a Soot plugin. Once PEPs are found, it extracts the corresponding permission(s).
This plugin performs an intra-method analysis and manages the following scenarios:
either (1) the permission is directly given as parameter, 
or (2) the permission value is initialized in a variable which is given as a parameter, 
or (3) an array is initialized with several permissions and is given as a parameter.
In every case we do a backward analysis of the method's bytecode using Soot's Unit Graphs which describe relations among statements of a
method. In the case where only one permission is given to the method, the first statement in the unit graph containing a reference to a valid Android permission String is extracted and the permission added to the list of the permissions needed by the method under analysis. In case of an array, all permissions of references to Android permission Strings are added to the list.

When no valid permission String is found, methods in the call-stack of the PEP method are analyzed. Indeed, permission String can be assigned indirectly to PEP methods.

\subsection{Handling Binder-based Communication}
\label{sec:binder}
Static analysis can not resolve call to services since they are done dynamically through the binder (see \ref{android}).
Since the binding uses a lookup table that is instantiated once at boot time within the \emph{system server}, our solution is to intercepted this lookup table and use it in a Soot plugin to redirect every proxy call to the concrete instance of the class which implements the service. 
In other terms, we feed the call graph engine with this domain specific information that it does not know from code.

Note that when using a field-sensitive (such as Spark) or context-sensitive analysis, services must be properly initialized. Otherwise, their fields would point to null and
method calls on those fields would not be considered during the call graph construction. We resolve this issue by providing a special initialization class to Spark containing services objects towards which remote service calls are redirected.

\subsection{Service Identity Inversion}
\label{sec:binder-identity}

In Android, services can call other services either with the identity of the initial caller (by default) or with the identity of the service itself. In the later case, remote calls are within \texttt{clearIdentity()} and \texttt{restoreIdentity()} method calls. When using the service identity, the permission checks are not done against the caller's declared permissions, but against the service's declared permissions. 
Since our goal is to compute the permission gap of an application (and not of system services), we can safely discard all permission checks that occur between calls to \texttt{clearIdentity()} and \texttt{restoreIdentity()}.
This significantly decreases the number of inferred permissions hence the number of false positives. 

For instance, let us assume that service S requires and declares permission $\theta$ which is not declared by application A. 
If A calls S, the code of S is executed  with the identity of A itself which would require A to declare $\theta$. 
To avoid this, the portion of code requiring $\theta$ is executed with S identity. 
Spark is flow-insensitive, so when we encounter calls to \texttt{clearIdentity()} or \texttt{restoreIdentity()}, we use an intra-procedural flow-sensitive analysis to discard permission checks that occur between those calls.

\subsection{Reflection in the Framework}
 
If the framework uses reflection, then the call graph construction is incomplete by construction.
Fortunately, the Android framework uses reflection in only 7 classes. We manually analyzed their source code.
Five of those classes are debugging classes. The \emph{View}
class uses reflection for handling animations. Finally, the \emph{VCardComposer} uses reflection in a branch that is only executed for testing purpose. 
In all cases, the code is not related to system resources hence no permission checks are done at all. This does not impact the static analysis of the Android framework.

\subsection{Dynamic Class Loading}

The Java language has the possibility to load classes dynamically. When used this features makes static analysis impossible since the loaded classes are only known at runtime. 
We found that eight classes of the Android system are using
the \texttt{loadClass} method. After manual check, six of them are system management classes and either are not linked to permission checks (ex: instrumenting an application)
or have to be accessed through a service.
Two are related to the \emph{webkit} packaged. They are used in the \texttt{LoadFile} and \texttt{PluginManager} classes. 
In both cases, permissions are checked \emph{before} loading classes, and not inside the loaded classes. Thus, there is no missed permission enforcement points either.

\subsection{Bytecode Manipulation Toolkits}
A last technical yet blocking issue was related to manipulation of Android bytecode.
We had to write the mapper and the sniffer on top of two different toolkits:
the mapper uses the Soot analysis framework developed at McGill University \cite{vall99soot};
the sniffer uses the ASM framework \cite{ASM}.
We had to use two different toolkits for the following reasons.
On the one hand, the code of the framework is open-source and written in Java, which is perfectly appropriate for an analysis using Soot.
On the other hand, since we do not assume to have the source code of end-user commercial applications, the application bytecode is only available as Dalvik bytecode.
While we can transform Dalvik bytecode to Java bytecode using a tool called ``ded'' developed at Penn State University\footnote{\url{http://siis.cse.psu.edu/ded/}}, the bytecode resulting from too many complex transformations is often not compatible with Soot for obscure reasons. The lower-level API of ASM enabled us to overcome these problems.

\subsection{Recapitulation}
We have presented the core technical issues we encountered while implementing our approach. We think that those problems may arise in other permission-based platforms than Android, and that identifying them and their solutions can be of great help for future work. Last not but not least, those points are crucial for replication of our results.

%% file: experimentations.tex
This section presents an evaluation of our approach.
First, we discuss the permission map extracted by static analysis using Soot and Spark.
Then we compare our results to the map extracted by Felt et al.\cite{Felt2011a} using runtime testing techniques.
Finally, we show that our approach actually detects permission gaps in real applications published in two different application stores.

\subsection{Extracted Permission Maps}

\input{table-soot-naive-vs-spark.tex}

In the Android v2.2 framework, 115 permissions are defined.
When predicting the required permissions of Android appplications, we want to guarantee that the inferred permission set is sound, i.e. that all inferred permissions are actually checked in code and that we do not miss some checks.
As said in Section \ref{subsec:def} (definition: Low Level Permission), and in Section \ref{subsec:service}, our static analysis method does not deal with: 
8 low-level kernel permissions; 
30 permissions checked at the level C++ services; 
8 permissions checked at the level of content provider.
Removing these permissions from the initial set of 115 permissions and by taking care of overlapping permissions (for instance, a permission can be checked at both C++ service and content provider levels) yields a set of \emph{71 high-level permissions}.   
In the following,  our discussion and comparison will only consider this set of 71 permissions.

For those 71 high level permissions, we claim that our static analysis at the framework level is sound.

We ran the static analysis based on Soot and which uses the
Spark call graph analysis described in \ref{sec:call-graph} augmented with
binder and service specifities (see Sections \ref{sec:binder} and \ref{sec:binder-identity}) on
the Android v2.2 framework bytecode. The resulting map is
summarized in table \ref{table:comparison-soot}. It gives the number of analyzed entry
points (methods of the framework), the number of methods
with no permission checks, the total number of permission
checks (ones in the matrix), and the number of methods
with at least one permission checks (with the median and
maximum number of permission checks).

According to this analysis, there are 9562 methods requiring
at least one permission, and among them, there are a median
of 2 permissions checked.

This fits with our developer experience with Android, the
methods have a clear scope and generally require a few
permissions (for instance, a method related to Bluetooth
management only requires permission BLUETOOTH). The
maximum of 50 permissions is related to methods which are
highly dependent of the usage context. In practice, developers only use the method indirectly and in a specific context
and declare a handful permission. However, from the blind
viewpoint of static analysis, there are 50 permissions involved in this method. There are only couple of such outlier
in the 9562 methods predicted by Spark to require at least
one permission. The question whether we are still sound,
i.e. whether we did not remove too many edges in the call
graph is answered in the next sub-sections.

The matrice is very sparse (it mostly contains zeros and a few ones -- the number of permission checks), because many methods do not contain permission checks and because one method checks at most an handful permissions.

In terms of CPU cost, the computation of the most CPU-intensive analysis, Soot Spark, is performed in about 11 hours on a Desktop Dell dual quad-core 2.4GHz with 24 Go RAM.

\subsection{Comparison with Felt et al.}
\label{sec:comparison-felt}

Let us now compare our results obtained with static analysis with the results of Felt et al.' obtained with testing ~\cite{Felt2011a}.
Both extract a list of required permissions for each method of the Android framework.
Felt et al.'s results contain 673 methods related with high-level permissions.
We analyze only 671 methods because 2 methods are related with application-specific objects provided in Felt's approach that are not available per construction in our static analysis approach.
Using our Spark-based static analysis approach with a maximum call graph depth of 10, for a given method, we either find the same permission set, or a larger one.
Our method never misses a permission that Felt et al. describe, this is piece of evidence of the soundness of our approach.

More precisely, we infer the same permission set per method signature for 552 methods (82.3\% of commonly analyzed methods). 
There is one ore additional permissions for 119 methods ( 1 additional permission for 118 methods, 2 for 1 methods).
There is no method for which we miss a permission,
Table \ref{table:comparison-felt} summarizes those results. Let us now discuss the discrepancy between our results.

\input{table-comparison-felt.tex}

The additional permissions are due to either analyzing irrelevant code or to missing input data in Felt et al.'s approach.
In the latter case, we are able to find permissions that are checked within specific contexts that were not taken into account by the generated tests of Felt et al. 
For instance, \texttt{MOUNT\-\_UNMOUNT\-\_FILESYSTEMS}  is only checked for me\-thod \texttt{Mount\-Service.\-shut\-down()} if the media (storage device) is ``\emph{present not mounted and shared via USB mass storage}'' (from the API documentation).
Another permission, \texttt{READ\-\_PHONE\-\_STATE} is needed for me\-thod \texttt{Cal\-ler\-In\-fo.get\-Cal\-ler\-Id\-()} only if the phone number passed in parameter is the 
voice mail number. Those test cases were not generated by Felt's testing approach. In real applications, test generation techniques can not guarantee a comprehensive exploration of the input space.

To us, these findings are typical when comparing a static analysis approach against a testing one: static analysis sometimes suffers from analyzing all code (including debugging and dead code, or code run in specific runtime environments), but is strong at abstracting over input data. 
On the other hand, testing must simulate as close as possible the production environment, but is cursed to always miss very specific usage scenarios.

Those results highlight the complementarity between static analysis and testing in the context of permission inference.
We think that the static analysis approach is complementary to the testing approach. Indeed, the testing approach yields an under-approximation 
which misses some permission checks whereas the static analysis approach yields an over-approximation in which those missing permission checks are found.
Using both approaches in collaboration would enable developers to obtain a lower and a upper bound of the permission gap. In particular, for an given Android applications, if both testing and static analysis approaches yield the same list of permissions, this list is the exact list of required permissions. This strong result is only possible by using both approaches in conjunction.

Note that we also compared the Spark based static analysis with a naive (Class Hierachy Analysis based) one which yields worst results (bigger permission sets). As expected, these
results show that the precision is higher when using Spark.

\subsection{Permission Gaps in Real Applications}
\label{sec:eval-ip}

We ran our tool on two datasets of Android applications. The first comes from an alternative Android Market\footnote{www.freewarelovers.com/android} and contains 1329 android applications.
For the second one, we consider the top 50 download applications of all 34 top-level categories of the Official Android Market, as well as the top 500 of all the applications and the top 500 of new applications (at the date of February, 23$^{rd}$ 2012).  As a result, after deduplicating the applications that appear in several rankings, the second dataset contains 2057 applications.
 
\textbf{Alternative Android Market:} 
For sake of soundness, we discard 587 applications that use reflection and/or class loading. Of the 742 remaining applications, 94 are declaring one or
more permissions which they do not use.
Consequently, we identify a permission gap for 94 Android applications.
We define the ``area of the attack surface'' with respect to permission gaps, as the number of unnecessary permission.
In all, among applications suffering from a permission gap, 
76.6\% have an attack surface of 1 permission,
19.2\% have an attack surface of 2 permissions, 2,1\% of 3 permissions and also 2,1\% of 4 permissions.

\textbf{Official Android Market:} 
For sake of soundness, we discard 1378 applications using reflection and/or class loading. On the 679 remaining applications, 124 are declaring one or
more permissions which they do not use.
In all, among applications suffering from a permission gap, 
64.5\% have an attack surface of 1 permission,
23.4\% have an attack surface of 2 permissions, 12.1\% of 3 or more permissions.

To sum up, those results show that permission gaps exists, and that our tool allows developers to fix the declared permission list in order to reduce the attack surface of permission-based software.

%% file: table-soot-naive-vs-spark.tex
 \begin{table}
\begin{tabularx}{\columnwidth}{p{4cm}|Y}
  
                           & SOOT Spark with binder\\
 \hline 
\# methods                 & 126660                           \\
 \hline 
\# permissions             & 71                               \\
 \hline 
\# methods with no check   & 112824                     \\
 \hline 
\# meth. with $\geq$ 1 perm. & 9562                        \\
~~median perm. checks      & 2                           \\
~~max perm. checks         & 50                          \\
 \hline 
\# perm. checks             & 137408                       \\
 
 \end{tabularx}
  \caption{Descriptive Statistics of The Permission Maps Found by Static Analysis}
   \label{table:comparison-soot}
 \end{table}

%% file: table-comparison-felt.tex
\begin{table}
\scriptsize
  \begin{center}
\begin{tabularx}{\columnwidth}{p{6cm}|X}
Permission set                  & Number of Methods\\
\hline 
\#Methods analyzed in \cite{Felt2011a}& 1282 \\
\#Methods with HL perm. only   & 673 \\
\hline 
Identical                       &  552 (82.3\%)\\
\hline 
 we find more permission checks & 119 (17.7\%)\\
 ~~~one more                    & 118 (17.6\%)\\
 ~~~two more                    & 1 (0.1\%)\\
\hline
[soundness evidence] we find less permission checks & 0 (0\%)\\
\end{tabularx}

  \caption{Comparison with Felt et al.~\cite{Felt2011a}. The discrepancy is due to the conceptual differences between static analysis and testing.}

  \label{table:comparison-felt}
  \end{center}
\end{table}

%% file: related-work.tex

We have presented an approach to reduce the attack surface of permission-based software.
The concept of ``attack surface'' was introduced by Manadhata and colleagues \cite{Manadhata2011}, it describes all manners \emph{in which an adversary can enter the system and potentially cause damage}. This paper describes a method to identify the attack surface of Android applications, which is a important research challenge given the sheer popularity of the Android platform.
In the context of Android, reducing the attack surface is adhering to the principle of least privileges introduced by Saltzer \cite{Saltzer1975}.

\subsection{On the Java Permission Model}

While the Android permission model is different from the one implemented in Java, the following pieces of research present related and relevant points to put our contribution in perspective.

Koved and al. described a new static analysis \cite{Koved:2002:ARA} to generate a permission list for a Java2 program (in the Java permission model).  An improved methodology was presented by Geay et al. \cite{conf/icse/GeayPTRD09}. We also use static analysis but in the context of Android which differs from a Java environment especially with respect to the binder mechanism linking Android API to services. As shown in our evaluation, the binder prevents off-the-shelf Java static analysis tools to resolve remote call to a service.\newline
%
%
Pistoia et al. \cite{conf/ecoop/PistoiaFKS05} presented a static analysis to identify portions of the
code which should be made privileged. 
This issue does not arise in the Android framework since code is not privileged per se, the access control is instead done at entry points. This means that the Android framework designers must be careful of creating unique entry points protected by permission enforcement points, but does not impact our static analysis.\newline
%
%
Role-based access control (RBAC) mechanisms are analyzed using static analysis by Centonze et al. \cite{CeNaFiPi2006}. 
When a protected operation manipulates data, this data should not be directly or indirectly accessible by a path not defined in the policy.  If not, the operation is said to be \emph{location-inconsistent}. 
The tool they developed can check  whether or not an RBAC policy for JavaEE programs is location consistent or present some flaws. The Android system defines permissions which
protects operation which in turn manipulate protected data. Our goal consists of computing permission gaps which may reveal a violation of the principle of least privilege. Whether Android protected operations are location consistent is out of scope of this paper.\newline
%
Also related to role-based access control, Pistoia et al. \cite{conf/icse/PistoiaFFY07} formally model RBAC and statically check the consistency of a JavaEE based RBAC system. We check that permission lists of Android applications respect the principle of least priviledge. 
The concepts are the same (Android permissions could be approximated to roles, and we check which roles are needed at every point of the Android framework) but the target systems are not. 
 Interestingly, we use a similar approach for solving the Binder problem as they do for solving the remote method invocation problem: instead of statically analyzing the Binder/RMI code which would not resolve the method, a mapping is computed from the call to a remote method to the remote method itself. A major difference though is that in the case of Android system services and context must be initialized beforehand to simulate a correct system state.

\subsection{On the Android Permission Model}

The Android security model has been described as much in the gray literature \cite{Enck2009,Shabtai2009} as in the official documentation \cite{android}.
Different kinds of issues have been studied such as social engineering attacks \cite{Hoffman2011}, collusion attacks \cite{Marforio2011}, privacy leaks \cite{Gibler2011} and privilege escalation attacks \cite{Felt2011b,Davi2011}.
In contrast, this paper does not describe a particular weakness but rather a software engineering approach to reduce potential vulnerabilities.\newline
%
However, we are not describing a new security model for Android as done by  \cite{Nauman2010,Ongtang2011,quire2011,Conti2011,Bugiel2011}. For instance, Quire \cite{quire2011} maintains at runtime the call chain and data provenance of requests to prevent certain kinds of attacks.
In this paper, we do not modify the existing Android security model and we devise an approach to mitigate its intrinsic problems.\newline
%
Also, different authors empirically explored the usage of the Android model.
For instance, Barrera et al. \cite{DBLP:conf/ccs/BarreraKOS10} presented an empirical study on how permissions are used. In particular, they used visualizing techniques such as self-organizing maps to identify patterns of permissions depending on the application domain, and patterns of permission grouping. Other empirical studies include Felt's one \cite{Felt2011} on the effectiveness of the permission model, and Roesner's one \cite{Roesner2011} on how users react to permission-based systems.
While our paper also contains an empirical part, it is also operational because we devise an operational software engineering approach to tame permission-based security models in general and Android's one in particular.\newline
%
Enck et al \cite{Enck2009a} presented an approach to detect dangerous permissions and malicious permission groups. They devised a language to express rules which are expressed by security experts. Rules that do not hold at installation time indicate a potential security problem hence a high attack surface. Our goal is different, we don't aim at identifying risks identified from experts, but to identify the gap between the application's permission specification and the actual usage of platform resources and services. Contrary to \cite{Enck2009a}, our approach is fully automated and does not involve an expert in the process.\newline
%
Finally, Felt et al. \cite{Felt2011a} concurrently worked on the same topic as this paper. They published a very first version of the map between developer's resources (e.g. API calls) and permissions. 
Interestingly, we took two completely different approaches to identify the map: while they use testing, we use static analysis. As a result, our work validates most of their results although we found several discrepancies that we discussed in details in Section \ref{empirical-study}. But the key difference is that our approach is fully automated while theirs requires manually providing testing ``seeds'' (such as input values). However, in the presence of reflection, their approach works better if the tests are appropriate. Hence, we consider that both approaches are complementary, both at the conceptual level for permission-based architectures, and concretely for reverse-engineering and documenting Android permissions.

%% file: conclusion.tex
In this paper, we have presented a generic approach to reduce the attack surface of permission-based software.
We have extensively discussed the problematic consequences of having more permissions than necessary and showed that the problem can be mitigated using static analysis.
The approach has been fully implemented for Android, a permission-based platform for mobile devices. 
Our prototype implementation is able to automatically find 9562 Android framework entry points which check permissions.
In a permission-based framework, all those checks have to be documented, hence our approach does a significant job in achieving this task in a systematic manner. 
For end-user applications, our evaluation revealed that 94/742 and 35/679 crawled applications from application stores for Android indeed suffer from permission gaps. 
We have also shown that our static analysis based approach is complementary to concurrent work \cite{Felt2011a} based on testing.

The security architecture of permission based software in general and Android in particular is complex. In this paper, we abstracted over several characteristics of the platform such as low-level permissions.
We are now working on a modular approach that would be able to analyze native code and bytecode in concert and to combine the permission information from both.
Furthermore, we are exploring how to express permission enforcement as a cross cutting concern, in order to automatically add or remove permission enforcement points at the level of application or the framework, according to a security specification.